\documentclass[english,aip,jcp,graphicx,twocolumn,showpacs]{revtex4}
\usepackage{ae}
\usepackage{aecompl}
\usepackage[T1]{fontenc}
\usepackage[latin1]{inputenc}
\usepackage{graphicx}

\makeatletter

\usepackage{float}

\usepackage{subfigure}

\makeatletter

\usepackage{float}

\usepackage{epsfig}

\makeatother

\usepackage{babel}
\makeatother
\begin{document}

\title{Localization and delocalization errors in density functional theory
and implications for band-gap prediction
\footnote[1]{Dedicated to
the memory of Professor Lorenzo Pueyo}}

\author{Paula Mori-S\'{a}nchez\footnote[4]{Both authors contributed equally to this work.}, Aron J. Cohen\footnotemark[4], and Weitao Yang }

\affiliation{Department of Chemistry, Duke University, Durham, North Carolina
27708}

\date{\today{}}

\begin{abstract}
The band-gap problem and other systematic failures of approximate
 functionals are explained from an analysis of
total energy for fractional charges. The deviation from the correct
intrinsic linear behavior in finite systems leads to delocalization
and localization errors in large or bulk systems. Functionals whose
energy is convex for fractional charges such as LDA display an
incorrect apparent linearity in the bulk limit, due to the
delocalization error. Concave functionals also have an incorrect
apparent linearity in the bulk calculation,  due to the localization
error and imposed symmetry. This resolves an important paradox and
opens the possibility to obtain accurate band-gaps from DFT.

\pacs{71.10.-w, 31.15.Ew, 71.15.Mb}
\end{abstract}
\maketitle

Accurate band-gap prediction is critical for applications in
condensed matter and nanotechnology. This unfortunately is not
currently possible within the widely used density functional theory
(DFT). The basic theory of the band-gap in DFT  has been addressed
 \cite{Perdew831884,Sham831888} but confusion remains. In our recent
 work\cite{Cohenxx}, it is shown possible to obtain the correct band-gap from
DFT calculations and demonstrate the theory for finite systems.

The main challenge is in the approximate exchange-correlation
functionals \cite{Becke883098,Lee88785,Perdew963865,Becke935648},
which, despite the success in a wide range of applications, still
suffer from systematic problems in describing charge-transfer
processes, excitation energies in molecules, response properties in
solids, electron transport and the band-gaps of semiconductors.
Previous understanding has focused on self-interaction-error and the
nature of the Kohn-Sham (KS) eigenvalues but recent work
\cite{Morisanchez06201102,Cohen07191109,Cohenxx,Ruzsinszky06104112}
has related these problems, more usefully, to the incorrect
description of systems with fractional charges.

In this Letter, we will resolve a paradox on the apparent linearity
of $E(N)$ for bulk systems for any approximate functionals and
 provide insight on the physical basis underlying
the systematic errors of functionals in large or extended systems,
and its implication in the calculation of the
 band-gap and other properties.

     The fundamental gap of an $N$-electron
semiconductor can be written as energy differences from
integer points \begin{eqnarray}
E_{{\rm gap}}^{{\rm integer}} & = & \Big\{ E(N-1)-E(N)\Big\}-\Big\{ E(N)-E(N+1)\Big\}\nonumber \\
 & = & I-A, \end{eqnarray}
 or as a difference of derivatives at $N$ \begin{equation}
E_{{\rm gap}}^{{\rm deriv}}=\left\{ \left.\frac{\partial E}{\partial N}\right|_{N+\delta}-\left.\frac{\partial E}{\partial N}\right|_{N-\delta}\right\} \label{derivgap}\end{equation}
 where $E_{{\rm gap}}^{{\rm integer}}=E_{{\rm gap}}^{{\rm deriv}}$
\emph{only if} the total energy is a straight line between the
integers, which is the case for the exact functional as shown by
Perdew \emph{et. al.} based on grand canonical ensembles
\cite{Perdew821691} and later by Yang \emph{et. al.} based on pure
states \cite{Yang005172}.

There  has been some question as to how the derivatives on the right
hand side of Eq. \ref{derivgap} are evaluated. In the case where
$E_{xc}$ is an explicit  functional of $\rho$, such as LDA or GGA,
then \begin{equation} E_{{\rm gap}}^{{\rm deriv}}=\epsilon_{{\rm
gap}}^{{\rm KS}}=\epsilon_{{\rm lumo}}-\epsilon_{{\rm
homo}}\end{equation}
 and $E_{{\rm gap}}^{{\rm deriv}}$ is just obtained from the KS eigenvalues. The detailed
expressions of $\left.\frac{\partial E}{\partial
N}\right|_{N\pm\delta}$ for other forms of exchange-correlation
functionals have been derived recently \cite{Cohenxx}.  When
$E_{xc}$ has an explicit
 dependence on the occupied KS orbitals, $E_{xc}[\phi_{i}]$,
 then  \cite{Cohenxx}
 \begin{equation} E_{{\rm gap}}^{{\rm
deriv}}=\epsilon_{{\rm gap}}^{{\rm KS}}+\left\{ \left.\frac{\partial
E_{xc}}{\partial N}\right|_{N+\delta}-\left.\frac{\partial
E_{xc}}{\partial N}\right|_{N-\delta}\right\}.
\label{OEP}\end{equation}
Thus, for such an orbital functional, the KS eigenvalues from an
optimized effective potential (OEP) \cite{Talman7636} calculation
are no longer the derivative of the energy expression. However, the
derivatives can be exactly the eigenvalues in a generalized
Kohn-Sham (GKS) calculation (e.g. Hartree-Fock calculations in the
case of exact exchange) \cite{Cohenxx}:
\begin{equation}
E_{{\rm gap}}^{{\rm deriv}}=\epsilon_{{\rm gap}}^{{\rm
GKS}}.\label{GKS}\end{equation} We see that the second term on the
right hand side of Eq. \ref{OEP}, which is labeled the derivative
discontinuity $\Delta_{xc}$ \cite{Perdew831884,Sham831888}, is
essentially the difference between an OEP and GKS calculation.
However, it does not contain the physics of the error of the
band-gap, as can clearly be seen from a consideration of the LDA or
GGA band-gap.

The error in the band-gap calculation using approximate functionals
is in the following
\begin{equation} E_{{\rm gap}}^{{\rm
integer}}=E_{{\rm gap}}^{{\rm deriv}}+\Delta_{{\rm
straight}},
\end{equation}
thus
\begin{equation} E_{{\rm gap}}^{{\rm
integer}}=\epsilon_{{\rm gap}}^{{\rm KS}}+\Delta, \end{equation}
where
\begin{equation}
\Delta=\Delta_{xc}+\Delta_{{\rm straight}}.
\end{equation}
Here $\Delta_{{\rm straight}}$, the difference between the gap from
finite difference and the derivatives,   accounts for the fact that
an approximate functional may not have the correct straight line
behavior between the integers. It is the consideration of this term,
completely missing in the literature, that is the key in
understanding the band-gap. We have seen that $\Delta_{{\rm
straight}}$ has an important effect in the energy gap in molecules
\cite{Cohenxx}. The focus of this Letter is to explore its
implications for extended systems.

\begin{figure}[!t]
\includegraphics[width=0.5\textwidth]{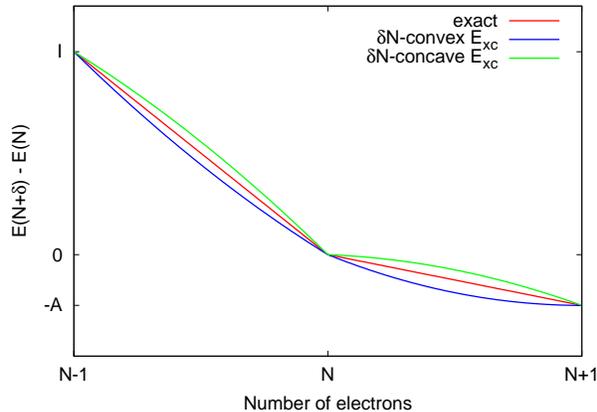}
\caption{\label{fig1} Energy vs Number of electron curve for a
finite system with three hypothetical functionals with: exact
straight line behavior, $\delta N$-convex behavior and $\delta
N$-concave behavior all which gave the same $I$ and $A$.}
\end{figure}

Now we present an important paradox for consideration of periodic or
bulk systems: It has been argued in extended systems that the
behavior of the total energy as a function of addition of an
electron must be a straight line \cite{Perdew831884,Sham831888},
therefore $\Delta_{{\rm straight}}=0$ for all functionals. How can
it be opposite to the case of the finite
systems\cite{Morisanchez06201102,Cohenxx}? Furthermore, for LDA,
$\Delta_{xc}=0$, meaning that $\Delta=0$.  What is then the origin
of the systematic error for LDA ?  It is well known
\cite{Onida02601} that the calculation of semiconductors and wide
gap insulators with a semi-local functional (such as LDA) has large
systematic errors, underpredicting the band-gap by up to several eV.

We will resolve this apparent paradox by understanding
more about the straight line behavior of energy functionals in solids.

\begin{figure}[!t]
\includegraphics[width=0.5\textwidth]{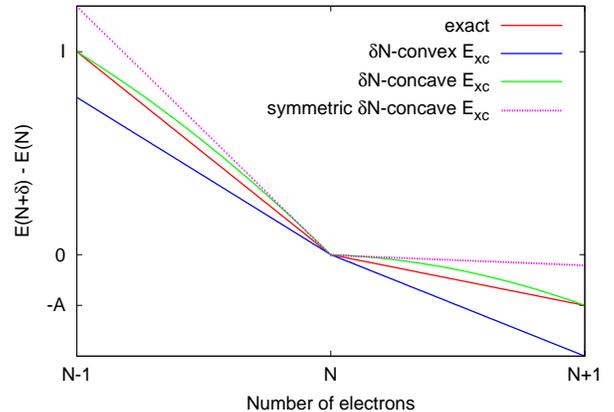}
\caption{\label{fig2} The same as Figure \ref{fig1} except taken to
the bulk limit. An additional curve is shown where crystal symmetry
is imposed on a $\delta N$-concave functional calculation.}
\end{figure}

The simplest periodic system to consider is a molecular crystal in
the infinite lattice constant limit. We start with a monomer unit
whose $E(N)$  curve for different functionals is as in Fig.
\ref{fig1}: the exact functional with correct straight line
behavior, one with incorrect convex behavior for fractional charges
($\delta N$-convex) such as LDA, and another with incorrect concave
behavior ($\delta N$-concave) such as HF theory.
Next consider adding an electron to more than one monomer unit. If
we first take the case of two monomer units, then we can see that
adding an electron to a $\delta N$-convex functional leads to half
an electron on each unit, as it is much lower than the energy of two
monomers, with $N$ and $N+1$ electrons respectively.
It is clear to see in this manner that with $M$ monomer units
\begin{equation}
ME(N+\frac{1}{M})<(M-1)E(N)+E(N+1)\label{concave}. \end{equation} As
the number of units  $M\rightarrow\infty$, the added $\delta$
electron delocalizes on to all the units such that the energy
approaches the initial slope of the $\delta N$-convex monomer curve
as in Figure 2:
 \begin{eqnarray}
E(MN+\delta) & = & ME(N+\frac{\delta}{M})\nonumber \\
 & \rightarrow & ME(N)+\delta \left.\frac{\partial E}{\partial N}\right|_{N+\delta}.\label{slope}\end{eqnarray}
In this way a functional like LDA, which is $\delta N$-convex for
small molecules, will have an \textit{apparent linearity} in large
or periodic systems. It is, however, a qualitatively incorrect
straight line, with the energy at the $N+1$ integer point dictated
by the fractional charge error of the functional. We can distinguish
this from the correct behavior of the exact functional, which has
\textit{intrinsic linearity} with correct integer points for all
$M$, whereas $\delta N$-convex functionals are only linear in the
limit $M\rightarrow\infty$ with incorrect integer points. This is
the \emph{delocalization error} of $\delta N$-convex functionals.

\begin{figure}[!t]
 \includegraphics[width=0.5\textwidth]{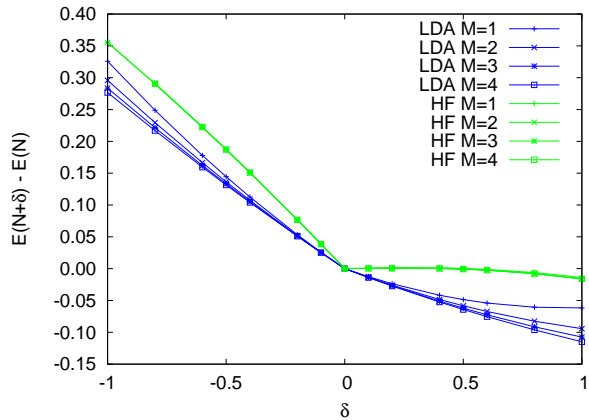}
\caption{\label{fig3} Energy for fractional numbers of electrons for
$({\rm H}_{16})_{M}$.}
\end{figure}

The situation is reversed for $\delta N$-concave functionals, where
the energy is always lower when the electron remains on one monomer
unit, even as $M$ increases. In fact the \textit{E vs. N} curve is
the same for all $M$. Delocalization actually raises the energy, as
the initial slope points to above the integer points such that the
inequality in Eq. (\ref{concave}) is reversed. For $\delta
N$-concave functionals we will find $\Delta_{{\rm straight}}\ne0$ if
we carry out an energy minimized calculation at $N+1$ in an infinite
system. This is the \emph{localization error} in $\delta N$-concave
functionals.
   However, if periodic symmetry is imposed
in the calculation, then the additional electron is delocalized
through the entire crystal as required by the translational
symmetry, a straight line will be seen that follows the initial
slope of the \textit{E vs. N} curve. This delocalized state also has
the energy of Eq. \ref{slope}, which is  a much higher energy,
imposed by the symmetry. All the above arguments apply both to the
addition of an electron and the addition of a hole (removal of
electron).

In a periodic system, we see the maximum effects of the localization
and delocalization error: for both $\delta N$-convex and $\delta
N$-concave functionals, we will have the apparent linearity for the
fractional charge, with the straight lines following the initial
derivatives and leading to too low band-gap for $\delta N$-convex
functionals and and too high band-gap for $\delta N$-concave
functionals. This explains the errors in the band-gap prediction.

In a finite system, \emph{delocalization error increases with system
size} until the apparent linearity appears. But localization error
stabilizes at certain system size. This also offers guidance on
calculations of band-gap for finite systems: The approach to
calculate the band-gap by explicit calculation of $I$ and $A$ from
subtraction and addition of electron to finite neutral species,
which works well with $\delta N$-convex functionals for small
molecules \cite{Cohenxx}, will not work for larger molecules,
because localization error increases leading to the incorrect nature
of the $N-1$ and $N+1$ points. However it may work for $\delta
N$-concave functionals, which do not suffer from the apparent
linearity problem (without translational symmetry) and may give
meaningful integer points. This issue is also slightly clouded by
the fact that HF theory may not give a reasonable energy for the
localized electron or even the right amount of localization. Hence
it will give an additional error to the integer points, but this has
a a different physical basis.

The discussion up till now has focused on the energy differences and
derivatives associated with the band-gap as it shows very clearly
the basic physical biases of approximate functionals. However, these
biases of the functionals have much wider implications. We can see
the differing behavior of the two types of functionals: $\delta
N$-convex (or LDA-type) functionals tend to delocalize electrons and
$\delta N$-concave (or HF-type) functionals tend to localize
electrons. So the nature of the electron density distribution is
dictated by the functional rather than the underlying physics of the
material. These  are the \textit{delocalization } and
\textit{localization }\textit{\emph{error}} of approximate
functionals. An intrinsically linear functional does not suffer from
these errors. The delocalization or localization error of
approximate functionals are responsible for much of the deviation of
calculated DFT properties from experimental results.
\begin{figure}[!t]
 \includegraphics[width=0.5\textwidth]{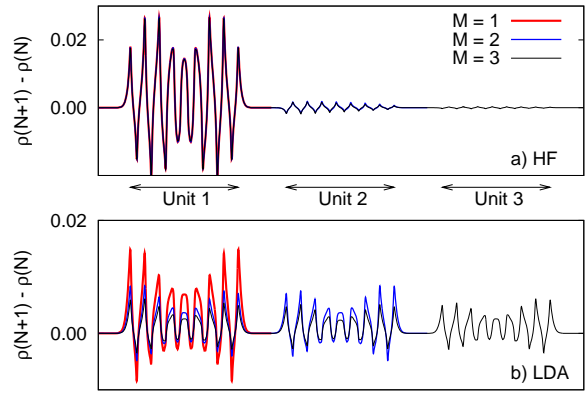}
\caption{\label{fig4} Density difference plot for $({\rm
H}_{16})_{M}$ system with HF on top and LDA bottom.}
\end{figure}

The idea of delocalization error introduced here is related to
many-electron self-interaction error (SIE)
\cite{Morisanchez06201102,Ruzsinszky06104112}. SIE has been blamed
for a poor description of the band-gap, but without an understanding
of the actual errors that it may introduce in to a particular
calculation. We believe that the terms localization and
delocalization error capture the physics of the problem in a more
useful manner than self-interaction error.

The argument above is carried out in the infinite lattice constant
limit but it is clear that same physical biases will be found at
finite lattice constant in a normal periodic calculation. To
illustrate the delocalization and localization errors we carry out a
simple calculation on a one-dimensional system based on previous
work \cite{Morisanchez0311001} of H$_{2}$ polymer polarizability. We
take a set of H$_{2}$ molecules that clearly shows the
characteristic behavior of Fig. 1, a chain made up of 16 atoms, and
repeat it with a 15 a.u. distance in between the units. The results
in Fig. 3 have the same behavior as in Fig. 2, showing that the
energy of a $\delta N$-concave functional (HF) remains the same,
independent of the number of units, $M$. With LDA we see the convex
behavior disappear as the delocalization error increase with M, and
the $E(N)$ curve becomes linear following the initial slope of the
monomer unit. We can see that the point at the integer corresponding
to the addition or subtraction is qualitatively wrong with a much
too low energy. Even changing the distance between the units from 15
a.u. to 5 a.u. has no effect on Fig. 3.

 Also in Fig. 4 we show a plot of the difference density
($\rho(N+1)-\rho(N)$) for different sized units. We observe a clear
difference between HF and LDA. HF localizes the extra electron to
just one of the units, whereas LDA delocalizes the extra electron
over all the units with a corresponding drop in energy. This clearly
shows the delocalizing bias of LDA and the systematic error it can
cause.

All these ideas tie in with the understanding of the band-gap
originally from Kohn \cite{Kohn64A171}, which relates the band-gap
to localization \cite{Resta99370}. A systematic error in the
band-gap implies a systematic error in describing localization,
which is just what we find. There will also be a systematic
qualitative error in the nature of the HOMO and LUMO orbitals and
their corresponding eigenvalues for large molecules, as well as the
conduction and valence bands for a crystal. These issues have
widespread relevance in the calculation of many electronic
properties of solids and large systems, as the delocalization error
in $\delta N$-convex functionals like LDA affects large molecules
and solids much more than small molecules.

The understanding given here also explains why TDLDA can be more
successful for the calculation of small molecules and metals than
for non-metallic infinite solids and polymers \cite{Onida02601},
where there is a basic problem in the description of the response of
the density due to the delocalization error. It can also explain why
Hartree-Fock and similar methods have problems with metallic systems
due to the fact the functional localizes electrons and therefore
opens a gap when the true nature should be to be delocalized with no
gap at the Fermi-level.

To conclude, we have shown that the  errors in the energy for
fractional charges in finite systems leads to systematic errors at
the integer points of larger systems. The addition of electrons or
holes is poorly described by $\delta N$-convex  functionals as they
delocalize the added particle. This leads to errors in the initial
slope as is seen by a band-gap calculation, but also means that the
explicit calculation of $I$ and $A$, for example in large cluster
calculations, will suffer from the same error. Functionals which
have opposite $\delta N$-concave behavior have the opposite tendency
 to localize electrons. This means that although there are
problems using the initial slope we can still gain information from
the explicit finite system calculation of $I$ and $A$.

The understanding offered in this work explains the physical nature
of the error in the band-gap from commonly used approximate
functionals, and shows the implications of this error to the
calculation of many other properties of solids, from optics to
electron transport. A path forward is shown:
by constructing  functionals free from localization or
delocalization error, one would be able to overcome most of the
problems.


Financial support from the National Science Foundation is greatly
appreciated. Discussions with Dr. San-Huang Ke and Dr. Xiangqian Hu have
been helpful.


\end{document}